\documentclass[10pt,aps,prb,twocolumn,superscriptaddress]{revtex4-1}

\usepackage{natbib}
\usepackage[dvipdfmx]{graphicx}
\usepackage{bm}
\usepackage{amsmath, amssymb}
\usepackage{upgreek}
\usepackage{color}

\newcommand{\BFA}{BaFe$_{2}$As$_{2}$}

\newcommand{\Tc}{$T_{\mathrm{c}}$}
\newcommand{\TAFO}{$T_{\mathrm{\mathrm{AFO}}}$}
\newcommand{\rc}{$\rho_c$}
\newcommand{\rcT}{$\rho_c(T)$}
\newcommand{\rab}{$\rho_{\mathit{ab}}$}
\newcommand{\rabT}{$\rho_{\mathit{ab}}(T)$}
\newcommand{\RRc}{RR$_{\mathit{c}}$}
\newcommand{\RRab}{RR$_{\mathit{ab}}$}
\newcommand{\cm}{cm$^{-1}$}
\newcommand{\plasma}{$\omega_{\mathrm{p}}$}
\newcommand{\wpsq}{$\omega_{\mathrm{p}}^{2}$}

\begin{document}

\preprint{ver.\ 4.1}

\title{Comprehensive study of out-of-plane transport properties in \BFA{}: Three-dimensional electronic state and effect of chemical substitution}

\author{M.~Nakajima}
\email[]{nakajima@phys.sci.osaka-u.ac.jp}
\affiliation{Department of Physics, Osaka University, Osaka 560-0043, Japan}
\author{M.~Nagafuchi}
\affiliation{Department of Physics, Osaka University, Osaka 560-0043, Japan}
\author{S.~Tajima}
\affiliation{Department of Physics, Osaka University, Osaka 560-0043, Japan}

%\date{\today}

\begin{abstract}

We investigated the out-of-plane transport properties of parent and chemically substituted \BFA{} for various types of substitution. Based on the studies of Hall coefficient and chemical-substitution effect, we have clarified the origin for the unusual temperature dependence of out-of-plane resistivity \rcT{} in the high-temperature paramagnetic-tetragonal phase. Electron (hole) carriers have an incoherent (coherent) character, which is responsible for non-metallic (metallic) \rcT{}. Although both of electron and hole contributions are almost comparable, a slightly larger contribution comes from electrons at high temperatures, while from holes at low temperatures, resulting in a maximum in \rcT{}. In the low-temperature antiferromagnetic-orthorhombic phase, the major effect of substitution is to increase the residual-resistivity component, as in the case for the in-plane transport. In particular, Co atoms substituted for Fe give rise to strong scattering with large \textit{ac} anisotropy. We found that K substitution induces a non-metallic behavior in \rcT{} at low temperatures, which is likely due to a weakly localized nature along the $c$-axis direction.

\end{abstract}

%\pacs{}
\maketitle

\section{Introduction}

High-transition-temperature (high-\Tc{}) superconductivity emerges in layered compounds such as copper oxides and iron pnictides. Copper oxides are characterized by a highly two-dimensional nature, which is definitely one of the key ingredients to give rise to high-\Tc{} superconductivity. In this two dimensional system, the electronic properties can be discussed only within the CuO$_2$ plane, while the out-of-plane charge dynamics does not play an important role in the essential physics. By contrast, it is known that two dimensionality or \textit{ac} anisotropy in iron pnictides is much smaller than in copper oxides, and thus the iron pnictides should be considered as an anisotropic three-dimensional system. Therefore, to comprehensively understand the electronic state of iron pnictides, it is indispensable to investigate not only the in-plane but also the out-of-plane properties. Nevertheless, there have been reported not many studies on the charge dynamics in the out-of-plane direction, partly because of the lack of large or thick single crystals.

\BFA{} family, the most extensively studied materials among iron pnictides, show a very small anisotropy. \cite{Tanatar2009,Tanatar2010} This is a multiband system, in which there are hole Fermi surfaces (FSs) at the center of the Brillouin zone and electron FSs at the zone corner. \cite{Yi2009,Yin2011} Since the hole FS is significantly warped around the Z point, the out-of-plane transport seems to be dominated by hole carriers. \BFA{} exhibits a magnetostructural phase transition from a paramagnetic-tetragonal (PT) to an antiferromagnetic-orthorhombic (AFO) phase at \TAFO{}, and superconductivity is achieved by suppressing the AFO phase with chemical substitution (or doping) or applying pressure. \cite{Ishida2009,Stewart2011} The AFO order has been attracted much interest because of its exotic electronic state associated with breaking of the fourfold rotational symmetry of lattice, spin, and orbital. \cite{Fernandes2014} The study of the in-plane resistivity using annealed single crystals revealed that the dominant role of chemical substitution in the AFO phase is to introduce disorder into the system and that a clear correlation is present between the magnitude of the residual resistivity (RR) and the in-plane resistivity anisotropy. \cite{Ishida2013a}

Out-of-plane transport properties of the \BFA{} family have been intensively studied by Tanatar \textit{et al.} \cite{Tanatar2009,Tanatar2010,Tanatar2011,Tanatar2013,Tanatar2014a,Tanatar2014b} They systematically measured the temperature dependence of the out-of-plane resistivity of as-grown \BFA{} substituted by various elements over a wide range of substitution. A focus was on the evolution of the resistivity in the PT phase with substitution, especially on a resistivity maximum at $T^*$, which was discussed in terms of pseudogap opening. \cite{Tanatar2010} This scenario seems reasonable but cannot account for a different composition dependence of $T^*$: Co substitution decreases $T^*$, \cite{Tanatar2010} whereas Ru, P, and K substitution increase it. \cite{Tanatar2013,Tanatar2014a,Tanatar2014b} As regards the effect of chemical substitution in the AFO phase, there has been no detailed study so far. A tendency of an increase in resistivity was seen in the results reported by Tanatar \textit{et al.}, \cite{Tanatar2010,Tanatar2013,Tanatar2014a,Tanatar2014b} implying that, as in the PT phase, the disorder effect associated with impurity scattering is also dominant in the AFO phase. A quantitative comparison with the in-plane transport properties is, however, difficult owing to a lack of precise out-of-plane resistivity measurements using annealed samples.

In this paper, we investigate the out-of-plane resistivity of parent and slightly substituted \BFA{}. To clarify the substitution effect also in the AFO phase, we chose the compositions showing the AFO order and used annealed single crystals for the measurement. We found that, in the PT phase of \BFA{}, contributions for the out-of-plane transport from electrons and holes, which have an incoherent and a coherent character, respectively, are comparable. Chemical substitution changes the balance between electrons and holes, and a resistivity maximum shows up at the temperature of dominant carrier crossover. In the AFO phase, the major effect of substitution for the out-of-plane transport is to introduce disorder. Co atoms are strong scatterers with large \textit{ac} anisotropy. K substitution induces an upturn of the resistivity at low temperatures, which would arise from a weakly-localized nature along the $c$-axis direction.

\section{Experimental}

Single crystals of parent and substituted \BFA{} were grown using the FeAs-flux method described elsewhere. \cite{Nakajima2010} For the growth of samples containing K, we used a stainless-steel container instead of a quartz tube to 
avoid a chemical reaction with vaporized K. \cite{Kihou2010} We prepared slightly substituted samples with various elements which show the AFO order: Ba(Fe$_{1-x}$Co$_x$)$_2$As$_2$, Ba(Fe$_{1-x}$Ru$_x$)$_2$As$_2$, BaFe$_2$(As$_{1-x}$P$_x$)$_2$, Ba$_{1-x}$K$_x$Fe$_2$As$_2$, and Ba$_{1-x}$Sr$_x$Fe$_2$As$_2$ (referred to as Co-Ba122, Ru-Ba122, P-Ba122, K-Ba122, and Sr-Ba122, respectively). Co and K substitutions work as electron and hole doping, respectively, and isovalent Ru, P, and Sr substitutions do not change the balance between electrons and holes. The substitution concentrations were determined by energy-dispersive x-ray analysis, and the $c$-axis lattice constant was in good agreement with the corresponding composition. The samples were annealed in an evacuated quartz tube together with BaAs at 800$^\circ$C for several days. \cite{Nakajima2011,Ishida2011}

The out-of-plane resistivity was obtained by the Montgomery method. \cite{Montgomery1971} The advantage of this method is that we can measure the out-of-plane and in-plane resistivity simultaneously. Samples were cut into rectangular solids with a dimension of (1.0--1.5) $\times$ (0.7--0.9) $\times$ (0.3--0.6) mm$^3$ ($a \times a \times c$ for high-temperature tetragonal axes) with care not to contain cracks or FeAs flux. Gold wires were attached to the four corners of the smaller \textit{ac} face with a silver paste (inset of Fig.\ 1). The length of 1.0--1.5 mm in depth is long enough to make almost no impact on the computed resistivity value. Measuring several times for each composition, we determined the absolute values of resistivity within the error of 10\%. At room temperature, the out-of-plane resistivity of the compositions investigated here ranges between 1.0 and 1.3 m$\Omega$\ cm. The out-of-plane Hall effect and magnetoresistance were measured on the sample of $0.5 \times 0.2 \times 1.0$ mm$^3$ ($a \times a \times c$) with the electrical current along the $c$ axis and the magnetic field perpendicular to the current. The resistivity value obtained in the magnetoresistance measurement using the four-probe method is consistent with that measured by the Montgomery method.

\section{Results}

\subsection{Transport properties of \BFA{}}

\begin{figure}
	\includegraphics[width=70mm]{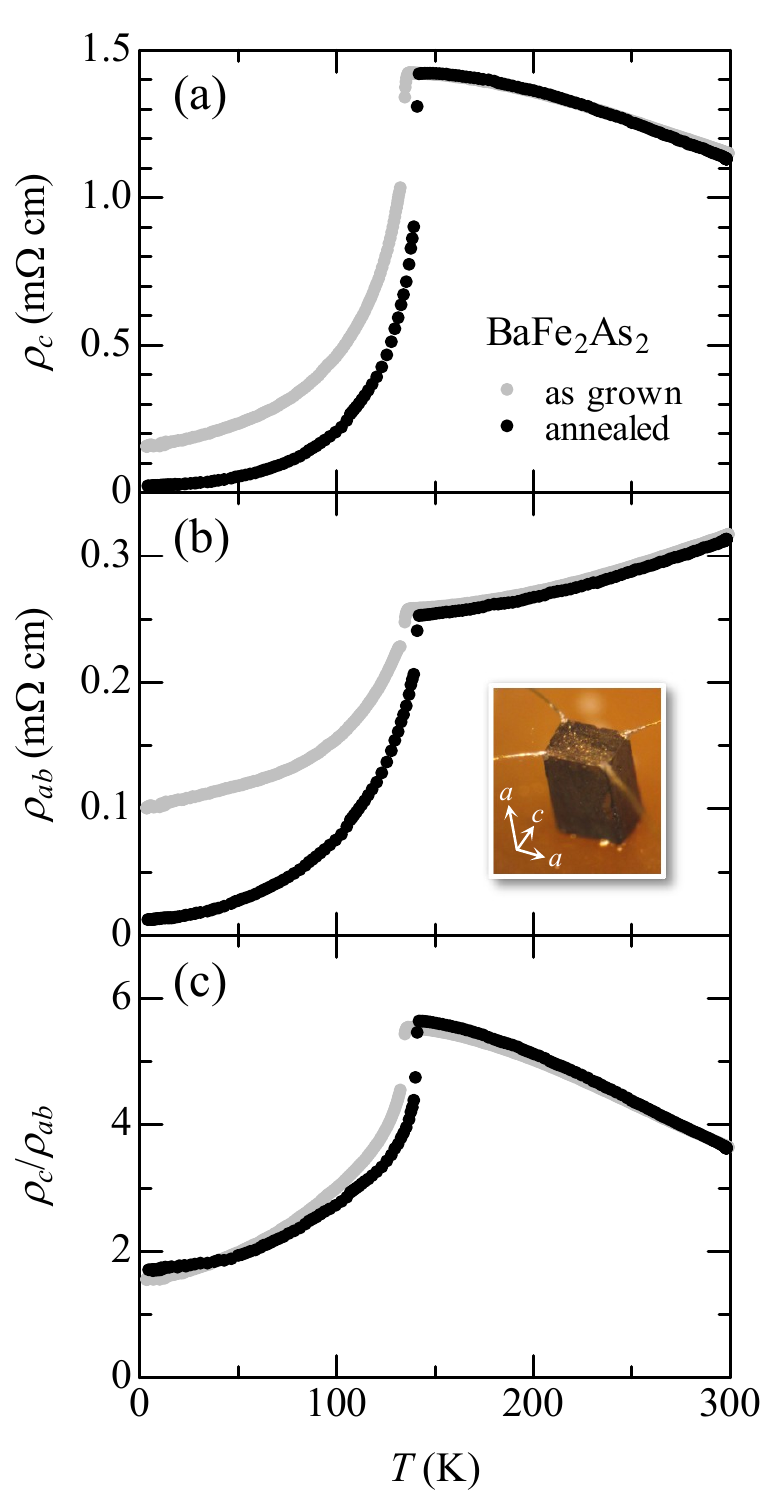}%
	\caption{Temperature dependence of (a) out-of-plane and (b) in-plane resistivity of as-grown and annealed \BFA{}. Inset shows a photograph of the sample for the Montgomery method. (c) \textit{ac} resistivity anisotropy \rc{}/\rab{} as a function of temperature.}
\end{figure}

Figures 1(a) and 1(b) show the temperature dependences of out-of-plane and in-plane resistivity (\rcT{} and \rabT{}) of \BFA{}, respectively. In the high-temperature PT phase, \rabT{} decreases with decreasing temperature, while \rcT{} monotonically increases down to the AFO transition temperature \TAFO{}. The behavior of \rcT{} is slightly different from the result reported by Tanatar \textit{et al.}, \cite{Tanatar2010,Tanatar2011,Tanatar2013,Tanatar2014a,Tanatar2014b} in which \rcT{} of \BFA{} shows a peak at $\sim$200 K and a metallic temperature dependence below the temperature. This might arise from admixture of an in-plane component because they measured slab-like samples with the shortest edge along the $c$ axis. Below \TAFO{}, both \rab{} and \rc{} show a significant decrease. The annealing treatment increases \TAFO{} from 137 K to 142 K and significantly reduces the resistivity in the AFO phase, whereas the resistivity in the PT phase does not show an appreciable change. The RR is quite small ($\sim$10 $\upmu \Omega$\ cm) both for \rc{} and \rab{}, indicating the high quality of the annealed sample.

The temperature dependence of the \textit{ac} anisotropy of resistivity \rc{}/\rab{} is shown in Fig.\ 1(c). The anisotropy is $\sim$4 at room temperature, increases with decreasing temperature, and takes a maximum value of $\sim$6 at \TAFO{}. Below \TAFO{}, the anisotropy turns to decrease and becomes less than 2 at low temperatures. Interestingly, no significant change in the anisotropy was observed between the as-grown and annealed samples. This indicates that the rates of the suppression of \rc{} and \rab{} by annealing are almost the same.

\begin{figure}
	\includegraphics[width=80mm]{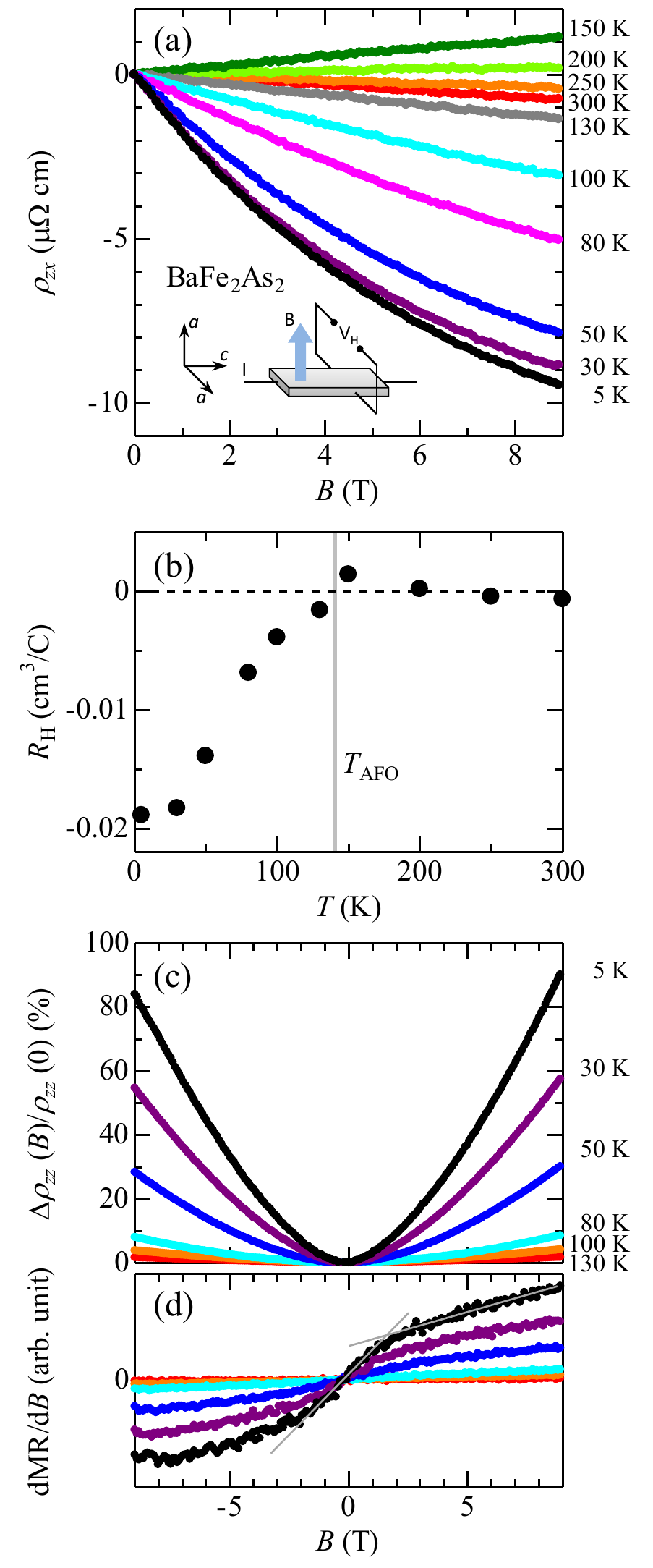}%
	\caption{(a) Magnetic-field dependence of the out-of-plane Hall resistivity $\rho_{\mathit{zx}}$ for annealed \BFA{}. Inset shows the configuration of the Hall-effect measurement. The sample was cut into bars with the longest edge along the $c$ axis. Current flows along the $c$-axis direction. (b) Out-of-plane Hall coefficient $R_{\mathrm{H}}$ as a function of temperature. $R_{\mathrm{H}}$ is determined from the slope of $\rho_{\mathit{zx}}(B)$ in the low-field limit. \TAFO{} is indicated by the vertical line. (c) Magnetic-field dependence and (d) the field derivative of the out-of-plane magnetoresistance for annealed \BFA{} at various temperatures below \TAFO{}. The gray lines indicate that the derivative of the magnetoresistance changes its slope at $\sim$1.5 T.}
\end{figure}

Figure 2(a) shows the magnetic-field dependence of the out-of-plane Hall resistivity $\rho_{\mathit{zx}}$ measured on annealed \BFA{} up to 9 T. $\rho_{\mathit{zx}}$ exhibits a linear field dependence above \TAFO{}, whereas the curve becomes concave upon entering the AFO phase. This behavior arises from the multiband nature with carriers with very different characters and is also observed for in-plane Hall resistivity of \BFA{}, \cite{Ishida2011} although the curvature is milder for the present out-of-plane case. In Fig.\ 2(b), the out-of-plane Hall coefficient derived from the data in Fig.\ 2(a) is plotted as a function of temperature. The Hall coefficient is negative at room temperature and increases with decreasing temperature, resulting in a sign change to positive below 200 K. Below \TAFO{}, the Hall coefficient turns to be negative again. The temperature dependence below \TAFO{} is monotonic but shows a shoulder structure at $\sim$100 K. At this temperature, a similar behavior is observed in the study of the in-plane Hall effect. \cite{Albenque2009} The present result indicates that the band-structure modification occurs three dimensionally in \BFA{}.

In Figs.\ 2(c) and 2(d), we show the magnetic-field dependence of out-of-plane magnetoresistance and its derivative with respect to magnetic field for annealed \BFA{}, respectively. The magnetoresistance is greatly enhanced with decreasing temperature in the AFO phase, whereas that is tiny in the PT phase (not shown). The magnitude reaches $\sim$90\% at $T$ = 5 K and $B$ = 9 T, indicative of quite high mobility of carriers, although the value is smaller than the in-plane magnetoresistance. \cite{Ishida2011} A kink structure observed in the derivative at $\sim$1.5 T [Fig.\ 2(d)] evidences the presence of $B$-linear component. This structure has been already reported in the in-plane magnetoresistance and is discussed in terms of a quantum transport of Dirac-cone states. \cite{Ishida2011,Huynh2011} Our result indicates that a Dirac nature is also present for the out-of-plane transport in \BFA{}, suggesting a three-dimensional Dirac system.

\subsection{Evolution with chemical substitution}

\begin{figure*}
	\includegraphics[width=170mm]{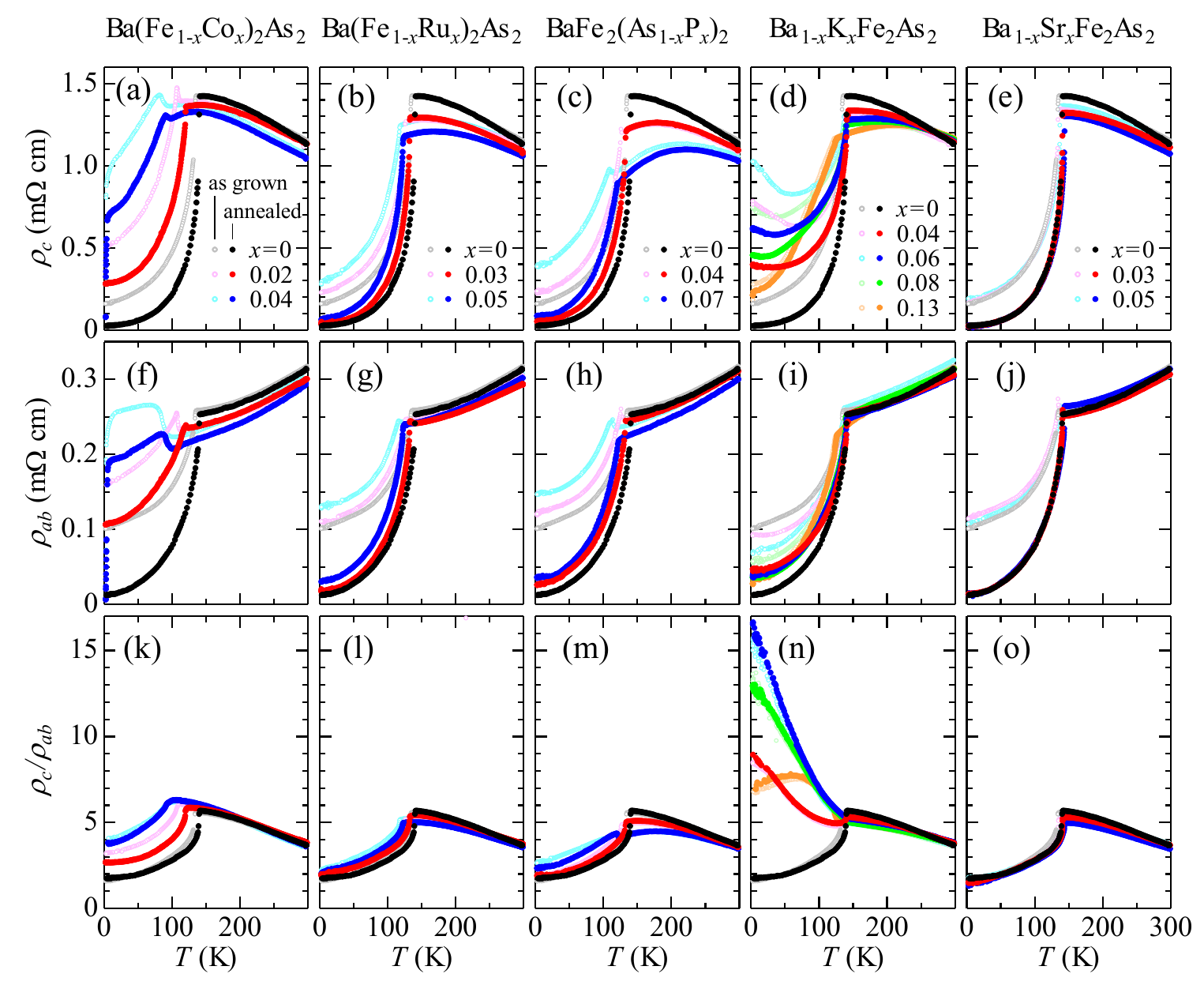}%
	\caption{Temperature dependence of resistivity for as-grown and annealed Co-, Ru-, P-, K-, and Sr-Ba122. (a--e) Out-of-plane resistivity \rcT{}, (f--j) in-plane resistivity \rabT{}, and (k--o) \textit{ac} anisotropy of resistivity \rc{}/\rab{}. Results for as-grown samples are indicated by open symbols.}
\end{figure*}

Figure 3 summarizes the results of the resistivity measurement for as-grown and annealed Co-, Ru-, P-, K-, and Sr-Ba122. The annealing effect for the substituted samples is the same as the case for the parent compound \BFA{}. The annealing treatment increases \TAFO{} by several kelvins. The resistivity in the PT phase is hardly affected by annealing, whereas that in the AFO phase dramatically decreases. The resistivity anisotropy does not change by annealing for the whole temperature range. Hereafter, we mainly focus on the results for the annealed samples.

In the PT phase, the in-plane resistivity does not show a large change with substitution within the composition range in the present work [Figs.\ 3(f--j)]. Only a small decrease in \rab{} is observed for Co- and P-Ba122. On the contrary, a pronounced change is seen in the out-of-plane resistivity. For Ru-, P-, and K-Ba122 [Figs.\ 3(b--d)], chemical substitution decreases \rc{} and significantly increases the slope of \rcT{} [d\rcT{}/d$T$], giving rise to a local maximum at $T^*$. \rcT{} of Co-Ba122 shows a downward parallel shift [Fig.\ 3(a)]. A local maximum shows up due to a suppression of the AFO order with substitution rather than a change in the slope of \rcT{}. For Sr substitution, only a subtle change was observed for the compositions investigated here [Fig.\ 3(e)].

The substitution effect in the AFO phase is rich in diversity depending on the species of substituted elements. For \rabT{}, it has turned out that the predominant effect of substitution is to introduce disorder, leading to an increase in the RR component. \cite{Ishida2013a,Ishida2013b,Liu2015} A large increase is indeed observed for Co substitution [Fig.\ 3(f)], in agreement with the previous study. \cite{Ishida2013b} On the other hand, the effect is small for isovalent substitutions: Sr substitution shows no discernible change [Fig.\ 3(j)], and a small resistivity increase is seen for Ru and P substitution [Figs.\ 3(g) and 3(h)]. \rabT{} for K-Ba122 shows a nonmonotonic evolution with $x$ [Fig.\ 3(i)]. The residual resistivity initially increases with K substitution but turns to decrease with further substitution after taking a maximum at $x$ = 0.04.

The composition dependence of \rcT{} below \TAFO{} is similar to that of \rabT{} except for K substitution. As in the case for \rabT{}, a large increase in \rcT{} is observed for Co substitution [Fig.\ 3(a)]. Ru and P substitution slightly increase the resistivity, while Sr substitution gives no appreciable change. On the contrary, \rcT{} for K-Ba122 shows not only a significant increase but also a peculiar temperature dependence at low temperatures. A clear non-metallic behavior was observed below 40, 35, and 20 K for $x$ = 0.04, 0.06, and 0.08, respectively. This feature is more prominent for as-grown samples.

The temperature dependence of the resistivity anisotropy in the PT phase is similar to \rcT{} because of a less temperature dependence in \rabT{}. The anisotropy at 300 K (\rc{}/\rab{} $\sim$4) is nearly independent of compositions. The anisotropy above \TAFO{} barely changes with Co substitution, whereas the other four substitutions decrease it. In the AFO phase, the anisotropy increases with substitution, although Sr substitution gives rise to an indiscernible change. For K-Ba122, the anisotropy at low temperatures is extremely large due to the non-metallicity in \rcT{}. The most enhanced anisotropy was observed for $x$ = 0.06 (\rc{}/\rab{} $\sim$16 at $T$ = 5 K).

\section{Discussion}

\subsection{Electronic state in \BFA{}}

The temperature dependence of \rab{} for \BFA{} exhibits a metallic behavior in the PT phase. Since \BFA{} is a multiband system, contributions from hole and electron carriers have to be considered. For the in-plane conduction, it turned out that holes are strongly scattered incoherent carriers and the dominant contribution comes from coherent electrons. \cite{Albenque2009,Nakajima2014} The small number of electrons gives rise to the relatively high resistivity at room temperature ($\sim$0.3 m$\Omega$\ cm) and a moderate temperature dependence, or a bad-metallic behavior.

On the other hand, \rcT{} above \TAFO{} increases with decreasing temperature. The FS of \BFA{} is composed of warped cylinders, among which the hole FS around the Z point are most warped, and thus the dominant carrier along the $c$ axis is considered to be holes. \cite{Tanatar2009} An issue to be addressed is why \rcT{} exhibits a non-metallic behavior in \BFA{} in spite of rather three-dimensional FS and small anisotropy. Tanatar \textit{et al.}\ pointed out that the presence of the resistivity minimum observed for heavily Co-doped samples, as well as low anisotropy values, rules out a crossover between incoherent and coherent transport. \cite{Tanatar2010} Together with the fact that the temperature dependence of the $^{75}$As NMR Knight shift for Co-Ba122 shows a mild slope change around $T^*$ and becomes weaker below $T^*$, \cite{Ning2009} they claimed that the carriers (presumably holes) are affected by a charge-gap formation, leading to the non-metallic temperature dependence. \cite{Tanatar2010}

The present result of the Hall coefficient in Fig.\ 2(b) offers an alternative interpretation for the out-of-plane transport. In the PT phase, the contributions of electrons and holes are nearly balanced, and the prevailing carrier changes from electrons at high temperatures to holes below 200 K. The slope of \rcT{} becomes gentler with lowering temperature, and \rc{} is almost temperature independent near \TAFO{} [Fig.\ 1(a)]. This indicates that, contrary to the case for the in-plane transport, electron carriers are incoherent and would be affected by a charge-gap opening, while hole carriers dominant at low temperatures are coherent. The origin of the $c$-axis charge gap has to be clarified, but this is beyond the scope of this paper. The presence of the incoherent carriers governing the high-temperature out-of-plane transport is consistent with the result of the $c$-axis optical spectroscopy on \BFA{}. \cite{Chen2010} In the $c$-axis optical conductivity spectrum, no clear Drude response was observed at room temperature, and a small Drude term becomes appreciable with decreasing temperature. The Drude term would arise from the warped part of the hole sheet. It should be noted that our scenario is in line with the NMR result. \cite{Ning2009} The slope change in the temperature-dependent Knight shift for \BFA{} is present at $\sim$200 K. At this temperature, the Hall coefficient changes its sign [Fig.\ 2(b)], and, unlike the result by Tanatar \textit{et al.}, \cite{Tanatar2010} \rcT{} for \BFA{} does not show a maximum. The contribution from incoherent electrons becomes minor below 200 K, which would lead to the milder but finite temperature dependence of the Knight shift.

Upon entering the AFO phase, both \rab{} and \rc{} suddenly decrease. Although the FS reconstruction across the AFO transition leads to the loss of the carrier density, a decrease in the scattering rate surpasses this effect, giving rise to the reduction of the resistivity below \TAFO{}. \cite{Albenque2009} The negative values of the Hall coefficient indicate that electrons dominate the $c$-axis conduction in the AFO phase as in the in-plane case. We observed the dramatic decrease in the anisotropy below \TAFO{}. This three-dimensional nature coincides with the reconstruction of the FS. \cite{Yin2011} The anisotropy decreases with decreasing temperature and takes a minimum value of $\sim$1.7 in the residual-resistivity region, in good agreement with the value estimated using the plasma frequency \plasma{} and the scattering rate $1/\tau$ obtained by optical spectroscopy. \cite{Chen2010,Optical}

\subsection{Substitution effect in the PT phase}

In the PT phase, the substitution effect is stronger in \rcT{} than in \rabT{}. For the compositions studied in the present work, \rabT{} shows only a tiny change with substitution above \TAFO{}. On the other hand, \rcT{} exhibits a dramatic evolution by a few percent chemical substitution. In particular, composition dependence is prominent for Ru-, P-, and K-Ba122, in which the low-temperature resistivity significantly decreases, giving rise to a resistivity maximum at $T^*$. With increasing substitution content, $T^*$ is raised. No correlated feature was observed in \rabT{} around $T^*$ (Fig.\ 3). For Co and Sr substitution, the temperature dependence hardly changes, but a resistivity maximum appears in the Co-Ba122 associated with a suppression of \TAFO{}. As mentioned above, incoherent electron carriers dominate the out-of-plane conduction at high temperatures, and a crossover of dominant carriers from incoherent electrons to coherent holes results in the weakening of the non-metallicity. Hence, it is natural to consider that a further increase in the hole contribution leads to a metallic behavior in \rcT{} and thus a resistivity maximum.

Chemical substitution affects the electronic state in various ways. First, it changes the FS by injecting charge carriers or by varying the (local) crystal structure via chemical pressure. Second, substituted atoms introduce disorders into the system, leading to carrier scattering. Among these effects, the disorder effect is minor in the PT phase as evidenced by a decrease in resistivity by substitution, \cite{Nakajima2014} although this effect is dominant in the AFO phase, \cite{Ishida2013a} as will be discussed later. For isovalently substituted Ru-, P-, and Sr-Ba122, the dominant effect of substitution should be chemical pressure. In addition to chemical pressure, a change in the carrier number has to be considered for Co- and K-Ba122, corresponding to electron and hole doping, respectively.

For Ru- and P-Ba122, a small amount of substitution significantly increases $T^*$. This can be understood in terms of the change in the FS induced by chemical pressure. Since the warping of one of the hole sheets around the Z point is enhanced by Ru and P substitution, \cite{Ye2012,Xu2012} the hole contribution to the out-of-plane transport becomes stronger, and the metallic temperature dependence manifests itself at higher temperatures. However, it seems difficult to explain why Sr substitution barely changes \rcT{}. To address this issue, we have to see the change in the crystal structure with substitution. It is well known that the band structure of iron pnictides is very sensitive to the local crystal structure, the As-Fe-As bond angle $\alpha$ \cite{Lee2008} or the pnictogen height from the Fe plane $h_{\mathrm{Pn}}$. \cite{Mizuguchi2010} Both Ru and P substitution make $\alpha$ and $h_{\mathrm{Pn}}$ larger and lower, respectively, \cite{Rotter2010,Sharma2015} but $\sim$5\% Sr substitution leads to only a tiny change in $\alpha$ and $h_{\mathrm{Pn}}$. \cite{Kirshenbaum2012} Indeed, the band structure of \BFA{} does not change largely by Sr substitution, whereas P substitution gives an appreciable change. \cite{Rotter2010} Thus, the Sr concentration in the present study is too small to change the local crystal structure inducing a significant change in the band structure. Note that SrFe$_2$As$_2$ exhibits metallic \rcT{} in the PT phase and shows a smaller anisotropy than \BFA{}, \cite{Tanatar2009} consistent with the highly warped FS of SrFe$_2$As$_2$. \cite{Ma2010}

Next, as to Co- and K-Ba122, $T^*$ varies in opposite ways: $T^*$ decreases with Co substitution and increases with K substitution. Co (K) substitution makes $\alpha$ and $h_{\mathrm{Pn}}$ larger (smaller) and lower (higher), respectively, \cite{Drotziger2010,Rotter2009} which means that the variation of $T^*$ cannot be explained by the chemical-pressure effect. Therefore, carrier doping should be considered as a major effect. K substitution corresponding to hole doping increases the number of coherent hole carriers, which facilitates a metallic behavior of \rcT{} starting from higher temperatures. Meanwhile, Co substitution decreases the hole carrier number and hence lowers $T^*$. In Ref.~\onlinecite{Tanatar2010}, for heavily Co-substituted compounds, a metallic behavior was observed at high temperatures. Electrons with a rather incoherent character would gain coherence by a large amount of electron doping.

Summing up, all these results of chemical-substitution effects on $T^*$ turn out to be a strong support for our proposed model for the out-of-plane conduction in the previous subsection.

\subsection{Substitution effect in the AFO phase}

\begin{figure}
	\includegraphics[width=70mm]{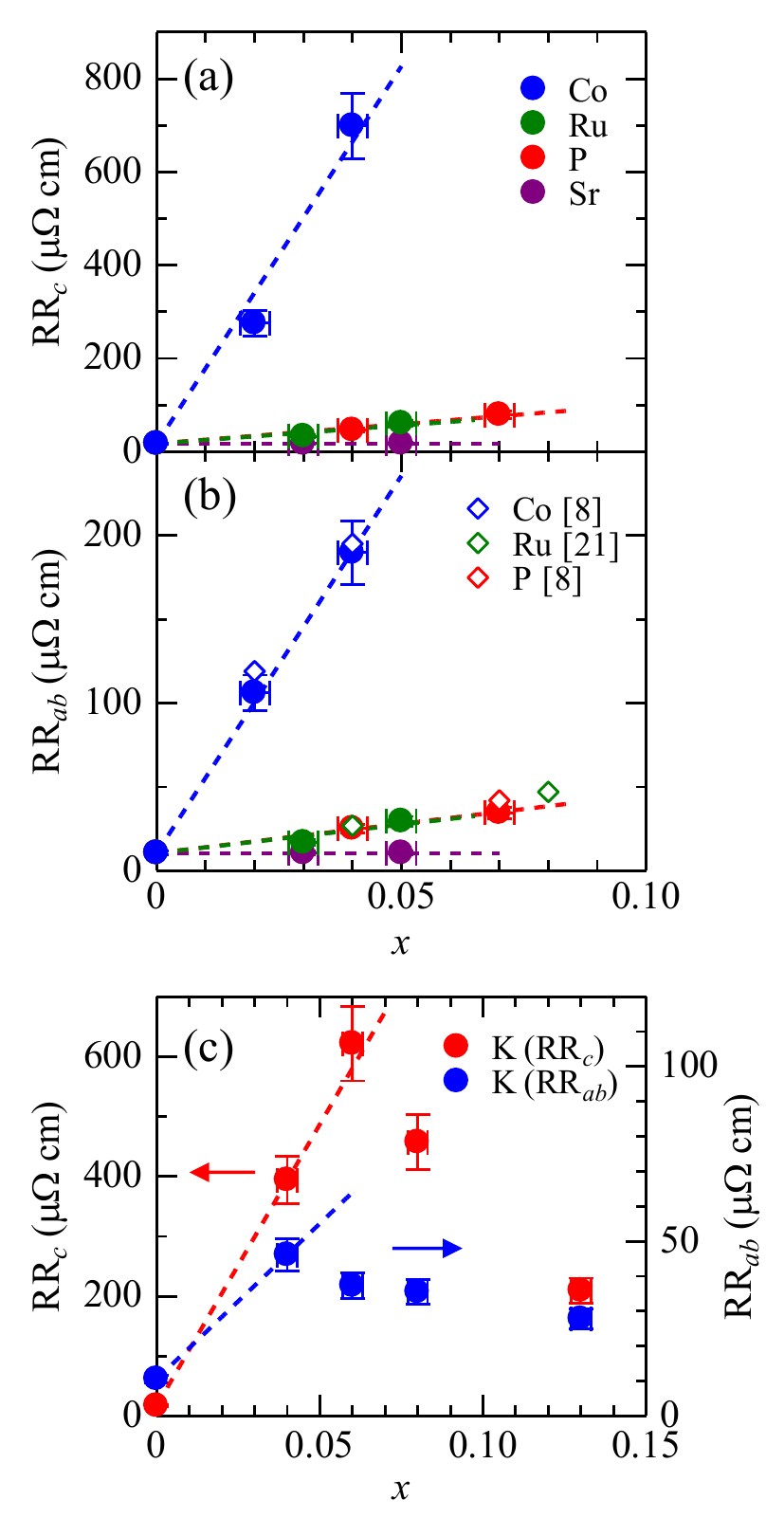}%
	\caption{Composition dependence of (a) out-of-plane and (b) in-plane RR for annealed Co-, Ru-, P-, and Sr-Ba122. The data of \RRab{} reported in Refs.\ \onlinecite{Ishida2013a} and \onlinecite{Liu2015} are also plotted. The \RRc{} and \RRab{} linearly increase with $x$ as indicated by the dashed lines. (c) Composition dependence of \RRc{} and \RRab{} for K-Ba122. In contrast to the other substitutions, the evolution with substitution is non-monotonic. Dashed lines just connect with the values for $x$ = 0 and 0.04.}
\end{figure}

Contrary to the evolution of the resistivity in the PT phase, chemical substitution increases \rc{} and \rab{} below \TAFO{}. Except for K and Sr substitution, both \rcT{} and \rabT{} shows almost a parallel upward shift, i.e., an increase in the residual-resistivity component. This indicates that the disorder effect dominates in the AFO phase. Such a prominent disorder effect stems from a small number of carriers enjoying exceedingly small scattering.

\begin{table}
	\caption{Increasing rate of out-of-plane and in-plane residual resistivity for the five substitution systems, corresponding to the slopes of the dashed lines shown in Fig.\ 4. The anisotropy of the rate is also shown. For K substitution, the RR does not show a monotonic composition dependence, and the values are calculated using \RRc{} and \RRab{} for $x$ = 0 and 0.04. No appreciable increase in RR was observed for Sr-Ba122.\vspace{1mm}}
	\begin{tabular}{p{31mm}p{9mm}p{9mm}p{9mm}p{9mm}p{9mm}}
		\hline \hline
		Element & \hfil Co \hfil & \hfil Ru \hfil & \hfil P \hfil & \hfil K \hfil & \hfil Sr \hfil \rule[0mm]{0mm}{4mm} \\ \hline
		\RRc{}/$x$ ($\upmu \Omega$ cm/\%) & \hfil 162 \hfil & \hfil 7.8 \hfil & \hfil 8.4 \hfil & \hfil 94$^{\mathrm{a}}$ \hfil & \hfil $<1$ \hfil \rule[0mm]{0mm}{4mm} \\
		\RRab{}/$x$ ($\upmu \Omega$ cm/\%) & \hfil 45 \hfil & \hfil 3.4 \hfil & \hfil 3.5 \hfil & \hfil 8.9$^{\mathrm{a}}$ \hfil & \hfil $<1$ \hfil \rule[0mm]{0mm}{4mm} \\
		Anisotropy & \hfil 3.6 \hfil & \hfil 2.3 \hfil & \hfil 2.4 \hfil & \hfil 11 \hfil & \hfil --- \hfil \rule[0mm]{0mm}{4mm} \\ \hline \hline
		\multicolumn{6}{l}{$^{\mathrm{a}}$The RR does not linearly increase with $x$. \rule[0mm]{0mm}{4mm}}
	\end{tabular}
\end{table}

To analyze the disorder effect in detail, we plot the RR of annealed samples as a function of substitution concentration $x$ in Fig.\ 4. The values were obtained from fitting of \rcT{} and \rabT{} below 50 K using a quadratic function of temperature. In Fig.\ 4(b), the in-plane RR (\RRab{}) for Co-, Ru-, and P-Ba122 reported in the previous studies are also plotted, \cite{Ishida2013a,Liu2015} in good agreement with the present result. It has already been clarified that \RRab{} linearly increases with substitution. \cite{Ishida2013a,Liu2015} In Fig.\ 4(a), a linear composition dependence is clearly seen for the out-of-plane RR (\RRc{}), evidencing again that the primary effect of chemical substitution in the AFO phase of \BFA{} is to introduce disorder into the system. In Table I, we show an increasing rate of RR for each chemical substitution. This value indicates the slope of the dashed lines in Fig.\ 4 and can be a good measure to see the strength of the disorder effect quantitatively. Among the five kinds of substitutions studied here, Co substitution shows the largest values both for \RRc{} and \RRab{}, indicating that introduced Co atoms work as very strong scattering centers. The strengths of impurity scattering for Ru and P are comparable and weak. The scattering generated by Sr atoms is extremely weak, keeping the RR unchanged within our experimental precision.

The increasing rate of \RRc{} with substitution is larger than that of \RRab{}. In Table I, the anisotropy of the slope is also shown. Neglecting the carrier multiplicity in iron pnictides, the resistivity is simply expressed as the Drude formula: $\rho = m^* / n e^2 \tau$, where $m^*$ and $n$ stand for the effective mass and the carrier number, respectively. In the system with impurity scattering, $1/\tau$ is written by a sum of the scattering rate of a pristine material $1/\tau_0$ and that originating from impurity $1/\tau_{\mathrm{imp}}$,
\begin{equation*}
\rho = \frac{m^*}{n e^2} (\frac{1}{\tau_0} + \frac{1}{\tau_{\mathrm{imp}}}) \propto \frac{1}{\omega_{\mathrm{p}}^2} (\frac{1}{\tau_0} + \frac{1}{\tau_{\mathrm{imp}}}).
\end{equation*}
Thus, when the impurity scattering is isotropic, the anisotropy of the disorder effect is determined by the anisotropy of \wpsq{} ($\propto n/m^*$) or the effective mass $m^*$. From the optical study, \cite{Chen2010} the anisotropy of \wpsq{} can be estimated to be $\sim$1.7. Although the value is slightly smaller than the anisotropy for Ru and P substitution, the anisotropy of the effective mass is considered to be a major cause to produce the anisotropic scattering for isovalent substitution.

The anisotropic effective mass alone, however, cannot explain the large anisotropy for Co substitution. Therefore, another mechanism has to be considered. One candidate is an anisotropic impurity state induced by a substituted Co atom in the AFO phase. In this scenario, Co atoms polarize its electronic surroundings anisotropically, giving rise to an anisotropic scattering cross section. The scanning tunneling spectroscopy measurement observed such an anisotropic state, \cite{Allan2013} and this has been discussed as one of the possibilities to produce the in-plane resistivity anisotropy. \cite{Nakajima2012,Ishida2013b} A similar anisotropic scattering may exist in the in-plane and $c$-axis direction. Another candidate is orbital-selective carrier scattering. The FS of iron pnictides is composed of all the five Fe 3d orbitals, and carriers with the 3d$_{z^2}$ character should play a crucial role in the out-of-plane transport. If a Co atom scatters carriers with the 3d$_{z^2}$ character more strongly, such orbital-selective scattering results in a larger increasing rate of \RRc{}.

Different from the other four substitutions, K substitution gives the non-systematic evolution of the resistivity in the AFO phase. As shown in Fig.\ 4(c), \RRc{} rapidly increases from the parent compound, takes a maximum at $x$ = 0.06, and shows a decrease with further substitution. On the other hand, \RRab{} reaches a maximum at $x$ = 0.04 and then moderately decreases. There is no correlation between the composition dependences of \RRc{} and \RRab{}. Moreover, the non-metallic temperature dependence was observed for \rc{} [Fig.\ 3(d)]. Some mechanism must be at work in addition to a simple disorder effect, as will be discussed in the next subsection.

Using the RR value at $x$ = 0.04, we estimated the increasing rates of \RRc{} and \RRab{} to be 94 and 8.9 $\upmu \Omega$~cm/\%, respectively. Although a simple comparison is impossible due to the distinct temperature dependence of resistivity, these values are smaller than those obtained for Co substitution. Interestingly, the estimated anisotropy ($\sim$11) is tremendously large, compared with the other substitutions. This can be understood in terms of the substitution lattice site. K atoms are substituted for Ba, which locates far away from the conduction Fe layer. Introduction of impurity into the blocking layer barely affects the in-plane conduction, but for the out-of-plane conduction, a change in the environment of the blocking layer should be highly influential for carriers moving along the $c$-axis direction.

Such a highly anisotropic behavior is not observed for Sr substitution. This would be because isovalent substitution of Sr$^{2+}$ for Ba$^{2+}$ can generate only tiny scattering, whereas aliovalent substitution of K$^{+}$ has a huge impact on the motion of carriers. In this sense, the strong scattering for Co doping might arise from the difference of the valence of Co from +2 as suggested by the near-edge x-ray absorption fine structure measurement. \cite{Merz2016} It is also worth noting that, among the three isovalent substitutions, the in-plane scattering induced by P atoms is not intermediate in magnitude between those by Ru and Sr but comparable with that for Ru substitution. \cite{Liu2015} This is counterintuitive because Ru substitution, which introduces disorder in the conduction layer, is expected to cause the strongest scattering. The strengths of the out-of-plane scattering for P and Ru substitution are also comparable. The present result indicates that As plays a crucial role in the electrical conduction both parallel and perpendicular to the Fe layers, consistent with the x-ray absorption measurements. \cite{Merz2016,Baledent2015}

\subsection{Origin of a resistivity upturn in K-Ba122}

The evolution of resistivity in the AFO phase with K substitution is distinct from that for the other substitutions. In particular, the resistivity upturn at low temperatures cannot be explained by the normal disorder effect which only increases the RR component. This feature is observed for the widest temperature range in \rcT{} for $x$ = 0.04, and a similar behavior, albeit very weak, is also observed for \rabT{} in the same composition. For $x$ = 0.06 and 0.08, the upturn is observed only for \rcT{}. Hence, such a behavior shows up in a narrower doping range for \rabT{}. This is why this characteristic feature has not been pointed out, although a signature was discernible with a careful inspection of previously reported data. \cite{Shen2011,Ohgushi2012}

One of the possible origins of the resistivity upturn is the Kondo effect. The scattering of conduction electrons in metals due to dilute magnetic impurities gives rise to a logarithmic temperature dependence of resistivity and thus a resistivity minimum at a finite temperature. This effect should work isotropically, but the resistivity minimum temperature is different in \rcT{} and \rabT{} of K-Ba122. A non-metallic behavior shows up only in \rcT{} for $x$ = 0.06 and 0.08. Moreover, a K atom cannot be considered as a magnetic impurity. These facts rule out the Kondo effect as the origin of the resistivity upturn.

\begin{figure}
	\includegraphics[width=80mm]{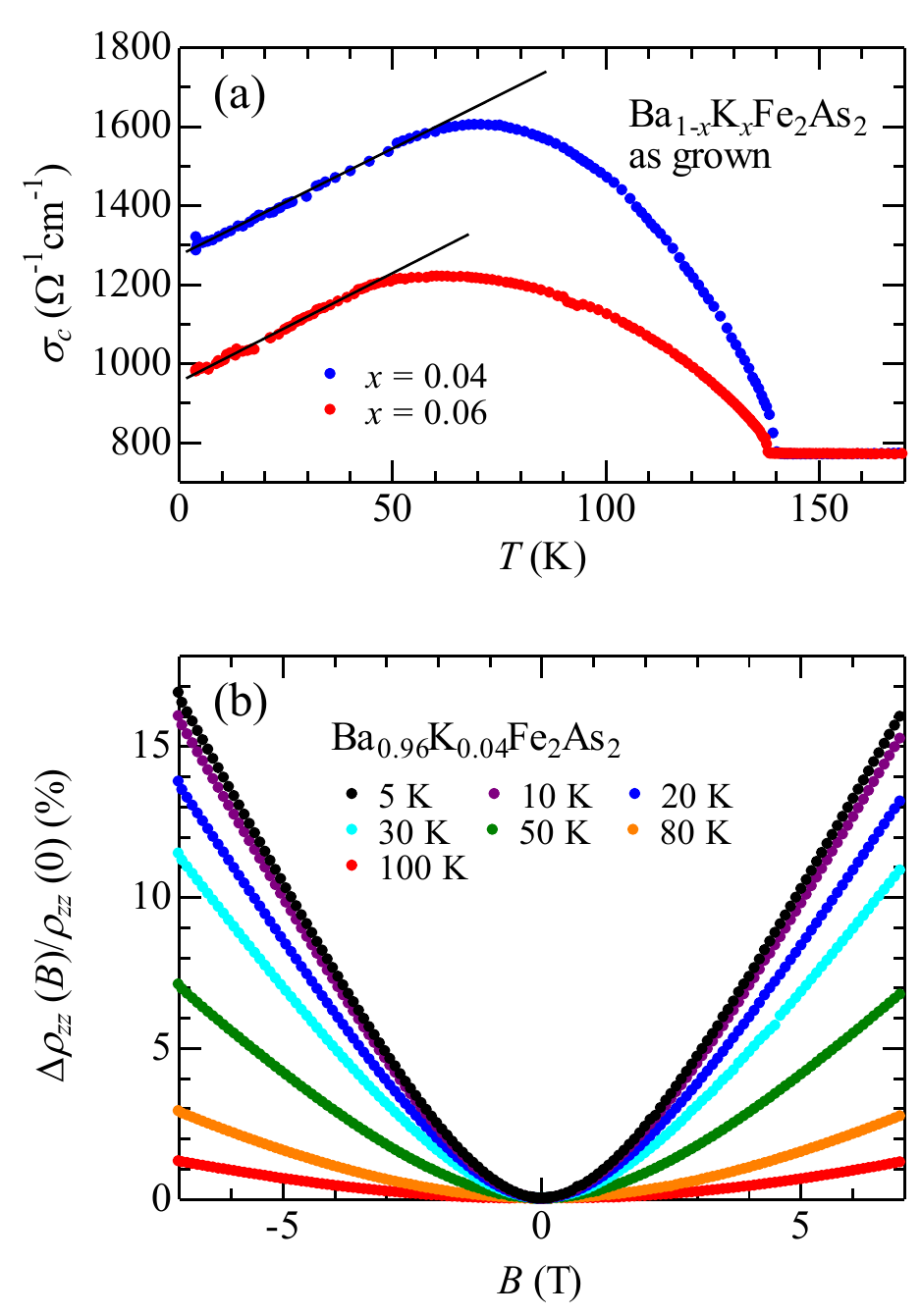}%
	\caption{(a) Temperature dependence of $c$-axis conductivity for as-grown K-Ba122 with $x$ = 0.04 and 0.06. The conductivity shows a $T$-linear behavior at low temperatures as indicated by the solid lines. (b) Magnetic-field dependence of the out-of-plane magnetoresistance for $x$ = 0.04.}
\end{figure}

A more plausible scenario is a weakly localized nature in a disordered system. In such a system, it is known that the conductivity shows a $\ln T$ and $T^{p/2}$ dependence for two and three dimensions, respectively, where $p$ is an exponent in the temperature dependence of the scattering rate $1/\tau \propto T^{p}$. \cite{Lee1985} Recently, resistivity measurements on iron-based ladder compounds Ba$_{1-x}$Cs$_x$Fe$_2$Se$_3$ under high pressure revealed that, the metallic samples show an insulating behavior at low temperatures. \cite{Hawai2017} The resistivity exhibits the $\ln T$ behavior, indicative of a weakly localized two-dimensional nature. In this compound, Ba atoms locating apart from the conducting layers are replaced with an alkali-metal element. This seems to be reminiscent of K-Ba122 in the present study. A pronounced upturn is observed in the as-grown crystals, and we show the temperature dependence of conductivity for $x$ = 0.04 and 0.06 in Fig.\ 5(a). For each composition, the conductivity exhibits a $T$-linear behavior at low temperatures. Applying to the three-dimensional case, we obtain $p \sim 2$, which implies that the inelastic scattering comes from the electron-electron interaction. The system is rather three dimensional, and only the carriers moving along the $c$-axis direction feel the random potential in the blocking layer induced by K substitution. The resistivity upturn vanishes for $x$ = 0.13 probably because the localization effect is suppressed by the change in the impurity potential and/or the carrier number.

One of the effective ways to verify the weak-localization effect is an observation of negative magnetoresistance. Figure 5(b) demonstrates magnetoresistance of as-grown K-Ba122 for $x$ = 0.04. Against expectations, we observed no signature of negative magnetoresistance. In disordered three-dimensional systems, the magnetic-field dependence of conductivity is proportional to $\sqrt{B}$, and its proportionality coefficient is independent of the details of the electronic structure of the system, \cite{Kawabata1980} allowing us to estimate the magnitude of negative magnetoresistance to be $\sim$0.6\% at 5 K and 7 T. Probably, the effect of weak localization is not strong enough to overcome rather large positive magnetoresistance resulting from the multiband nature. 

\section{Summary}

We performed the out-of-plane transport measurements on parent and slightly substituted \BFA{} with the five kinds of elements. In the PT phase of the parent compound, we revealed almost balanced contributions of electrons and holes to conduction along the $c$ axis. Contrary to the in-plane transport, the electrons moving along the $c$ axis are incoherent. With lowering temperature, the contribution from coherent holes grows and exceeds that from electrons at low temperatures. Chemical substitution changes the balance between holes and electrons via the local-structure change and/or the carrier-doping effect, and a crossover of the dominant carrier with temperature manifests itself as a maximum in \rcT{}. In the AFO phase, an introduced atom works predominantly as a scattering center. Each substituted atom generates a different strength of carrier scattering, resulting in a different rate of increasing resistivity with substitution. Co atoms substituted at the center of conduction plane are very strong scatterers with a large \textit{ac} anisotropy. Isovalent substitutions of Ru and P give rise to a moderate increase in the resistivity, and Sr substitution makes no effect. The resistivity upturn was observed prominently in \rcT{} of K-Ba122, which makes the resistivity for K-Ba122 highly anisotropic. The upturn is likely attributed to a weakly localized nature along the $c$ axis. Our result highlights the diversity of the substitution effect on \BFA{} depending on the lattice site and the element species.

\section*{Acknowledgments}

%\begin{acknowledgments}

The authors thank S.~Miyasaka for valuable discussions and M.~Ichimiya and T.~Nakano for their technical help. This work was supported by JSPS KAKENHI Grant Number JP26800187.

%\end{acknowledgments}

\end{document}